\documentclass[sigconf]{acmart}
\usepackage{url}
\usepackage[utf8]{inputenc}
\usepackage{caption}
\usepackage{graphicx}
\usepackage{subcaption}
\usepackage{amsmath}

\usepackage{mathtools,amsfonts,amsthm}  % mathtools loads amsmath
\usepackage{lipsum}
\usepackage{booktabs}
\usepackage{paralist}
\usepackage{xcolor,colortbl}
\usepackage{xspace}
\usepackage{booktabs}
\usepackage{color}
\usepackage[ruled]{algorithm2e}
\usepackage{multirow}
\usepackage[disable,colorinlistoftodos]{todonotes}

\newcommand{\vizai}{\textsf{VizAI}\xspace}
\SetKwInOut{Input}{Input}
\SetKwInOut{Output}{Output}

\copyrightyear{2022} 
\acmYear{2022} 
\setcopyright{acmlicensed}\acmConference[CODS-COMAD 2022]{5th Joint International Conference on Data Science \& Management of Data (9th ACM IKDD CODS and 27th COMAD)}{January 8--10, 2022}{Bangalore, India}
\acmBooktitle{5th Joint International Conference on Data Science \& Management of Data (9th ACM IKDD CODS and 27th COMAD) (CODS-COMAD 2022), January 8--10, 2022, Bangalore, India}
\acmPrice{15.00}
\acmDOI{10.1145/3493700.3493717}
\acmISBN{978-1-4503-8582-4/22/01}

\begin{document}
\title{VizAI : Selecting Accurate Visualizations of Numerical Data }

\author{Ritvik Vij}
\authornote{Four authors contributed equally to this research.}
\affiliation{
	\institution{Department of CSE, IIT Delhi}
	\country{India}
}
\email{Ritvik.Vij.cs517@cse.iitd.ac.in}
\author{Rohit Raj}
\authornotemark[1]
\affiliation{
	\institution{Department of CSE, IIT Delhi}
	\country{India}
}
\email{rohit.raj21@alumni.iitd.ac.in}
\author{Madhur Singhal}
\authornotemark[1]
\affiliation{
	\institution{Department of CSE, IIT Delhi}
	\country{India}
}
\email{msinghal1998@gmail.com}
\author{Manish Tanwar}
\authornotemark[1]
\affiliation{
	\institution{Department of CSE, IIT Delhi}
	\country{India}
}
\email{manish.tanwar@alumni.iitd.ac.in}
\author{Srikanta Bedathur}
\authornote{Contact author.}
\affiliation{
	\institution{Department of CSE, IIT Delhi}
	\country{India}
}
\email{srikanta@cse.iitd.ac.in}
% \copyrightyear{2019}
% \acmYear{2019}
% \setcopyright{acmlicensed}
% \acmConference{CIKM'19}{November 3--7, 2019}{Beijing, China}
% \acmPrice{15.00}
% \acmDOI{xx.xxxx/xxxxxxx.xxxxxxx}
% \acmISBN{xxx-x-xxxx-xxxx-x/xx/xx}
% \keywords{Visualizations; Machine Learning; HCI; Data Presentation}

% \settopmatter{printacmref=false, printfolios=false}

\begin{abstract}
A good data visualization is not only a distortion-free graphical representation of data but also a way to reveal underlying statistical properties of the data. Despite its common use across various stages of data analysis, selecting a good visualization often is a manual process involving many iterations. Recently there has been interest in reducing this effort by developing models that can recommend visualizations, but they are of limited use since they require large training samples (data and visualization pairs) and focus primarily on the design aspects rather than on assessing the effectiveness of the selected visualization. 

In this paper, we present \emph{VizAI}, a generative-discriminative framework that first generates various statistical properties of the data from a number of alternative visualizations of the data. It is linked to a discriminative model that selects the visualization that best matches the true statistics of the data being visualized. VizAI can easily be trained with minimal supervision and adapts to settings with varying degrees of supervision easily. Using crowd-sourced judgements and a large repository of publicly available visualizations, we demonstrate that VizAI outperforms the state of the art methods that learn to recommend visualizations. %are often visually appealing and statistically insightful about the data. 

%%Generating a good data visualization is essential not only for the ease of understanding but also to get insights about underlying statistical properties of the data. 
%% Despite its importance in various data analysis steps, visualizations are often prepared manually by experts. Recently there has been interest in reducing this manual effort by learning models that can recommend good visualizations for a given tabular data using carefully crafted battery of features. In this paper, we assert that a good visualization should (a) accurately convey the aggregate statistical information in the data, and (b) must capture salient features in the data. 
%We present an encoder-decoder framework, where the encoder stage consists of a deep convolutional neural (CNN) model trained to learn aggregate statistics of data from a given visualization. Subsequently we use use this to select the best plot for the data in an unsupervised fashion. %Quantifying the extent to which a visualization captures the salient features is done through a supervised (decoder) stage. 
% We propose VizAI, an end-to-end visualization recommendation system, that integrates the two stages allowing us to deal with domains with varying degrees of supervision. It can also reveal the best combinations of columns from a multi-columnar data to visualize based on their ability to reveal the underlying interestingness of the relation between columns. Using crowd-sourced judgments, we demonstrate that VizAI learns to predict the best plot for the data, comparable to manually created ones.
\end{abstract}
\maketitle

\section{Introduction}
% -- why do we need automation in visualization?
As more and more sectors of society ranging from government to businesses move to adopt data-enabled approach to decision making, it is not only imperative to have access to large volumes of data, but also to have tools that help get actionable insights from data as quickly as possible. Most users across these domains are not experts, as a result, effectively visualizing the data has become a crucial tool for obtaining insights. But generating effective visualizations is not an easy task, and requires not only a good understanding of the domain but also an expertise with the visualization tools. It has led to a growing interest in developing methods that can automate the process of visualizing datasets itself~\cite{webdb18,Dibia2018Data2VisAG,Luo2018DeepEyeTA,vizml}.

%With the advent of big data analytics and processing tools, institutions across all domains want to make use of data as efficiently as possible. Data visualization has become a crucial tool and the current tools rely mostly on manual specification through code or a user-interface. As a result, data visualization is available only for domain experts or people with expertise in using these tools. It is the need of the hour to have automated visualization which not only generates visually appealing plots but also conveys maximal information about the underlying data.

% -- what has been done so far, and why are they not adequate?
Current approaches taken towards this can be broadly classified into (a)~those using rule-based generation of visualizations~\cite{webdb18,Luo2018DeepEyeTA} and (b)~those based on machine models learnt on a large pool of existing visualizations~\cite{Dibia2018Data2VisAG,vizml}. Unfortunately, both these approaches suffer from serious limitations: on one hand, the rule-based methods have the obvious limitation that they can not easily adapt to new data or visualization paradigms. On the other hand, the machine learnt models rely on the availability of a large volume of prior data and its visualization to train on, making them hard to easily adapt to new visualizations. For example, the latest VizML system~\cite{vizml} uses more than 2.3 million data-visualization pairs to train, making it harder to replicate and highly dependent on the quality of visualizations chosen by the crowd. Apart from this, they also are affected by the (un)availability of certain visualizations in the training corpus. Further, neither approach considers explicitly quantifying how well the generated / recommended plot performs in terms it capturing the underlying data. %as a measure to select the best plots. They depend highly on the user's preference of some plot type to predict. 
%%%% This is written better in previous work.

\begin{figure}[t]
\begin{center}
   \includegraphics[width=1.\linewidth]{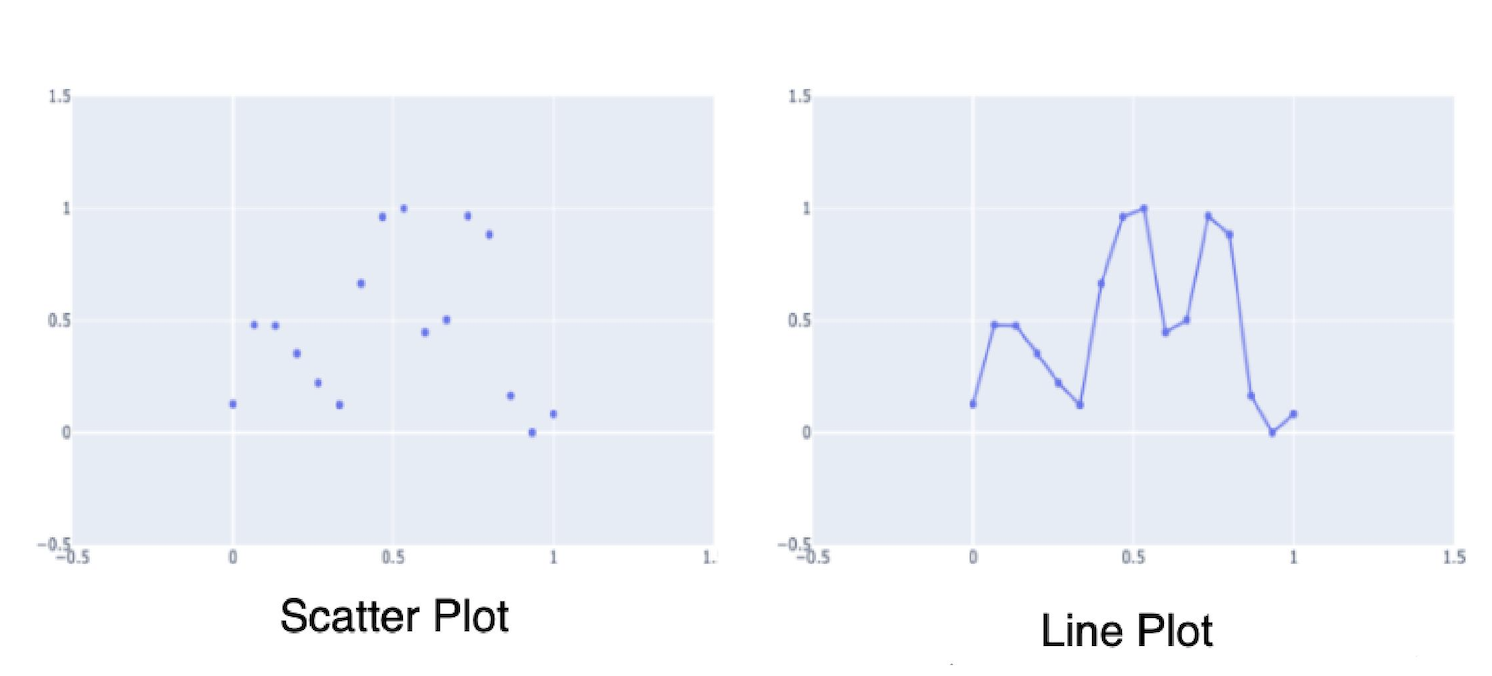}
   \includegraphics[width=0.47\linewidth]{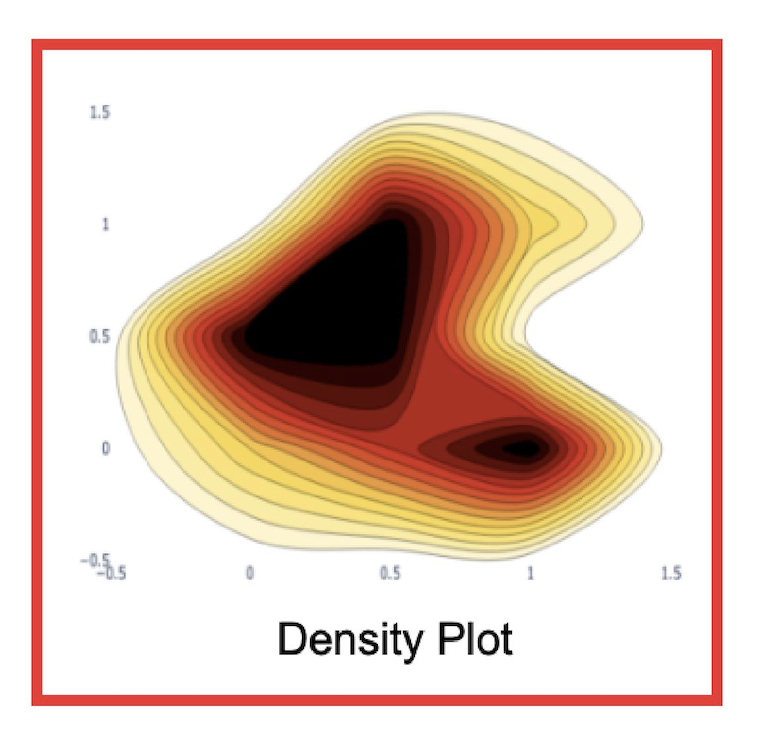}
\end{center}
\caption{Example of Visualization Recommendations by \vizai. All three visualizations represent the same data. \vizai selected the Density visualization (highlighted in red box), and in our crowdsourcing evaluation it had a perfect agreement among all judges as the best visualization choice.}
\label{fig:onecol}
\end{figure}

% -- how are we different and what are the advantages?
In this paper, we present \vizai\footnote{pronounced similar to Hindi word  \textit{vijay} (n): meaning \emph{success.}}, a machine learnt visualization recommendation framework that not only considers the preference of the user to a specific type(s) of chart, but, more importantly, also takes into account how well the specific visualization performs in \emph{representing} the underlying data accurately. In other words, \vizai explicitly builds on two of Tufte's canons for good visualization~\cite{tufte2001}, \emph{viz.,} 
\begin{inparaenum}[(a)]\item ``a good visualization should be a distortion-free representation of the underlying data'' and \item ``be closely integrated with the statistical description of the dataset''. \end{inparaenum} %We not only consider the user's preference of plot types for some particular style of data but also take in the fact how well that plot performs in representing the data. 
\vizai automatically generates a number of alternative visualizations for the given dataset, evaluates them on how well they (statistically) represent the underlying data and then selects the best performing one. The evaluation of the visualization itself is done using a deep convolutional neural network, based on an ensemble of standard ConvNets, that extracts features from the visualization which can be used to predict the aggregate statistics of the data. %and  For this evaluation we use a deep convolutional neural network model to extract features from the plot and use it to predict the aggregate statistics of the data.
For example, in Figure~\ref{fig:onecol} the underlying data was plotted using scatter, line and density plots (we used Chart Studio by Plotly for plotting these, although our work can be easily used with any other plotting software). Of these, \vizai ranked the \emph{density} plot as the best and this was further confirmed by our crowdsourcing evaluation (details are in Section~\ref{sec:crowd}).

% Note that our focus in this paper is on generating visualizations of numerical datasets rather than textual or text-heavy content. Often the textual data visualization involves transforming the text into some representation in a numerical domain (e.g., by sentiment extraction, word count/co-occurrence distribution, etc.), and \vizai can be easily adapted to those settings if desired. With numerical data, commonly used  visualizations are line, scatter, area, bar and pie charts, and more sophisticated settings require \emph{density} plots.

{Note that our focus in this paper is on generating visualizations of numerical datasets rather than datasets with varied data types, such as textual, categorical, temporal, etc., across data fields.  Often these data types are transformed as part of data preprocessing into an appropriate numerical representation before their visualization~\cite{Dibia2018Data2VisAG, Luo2018DeepEyeTA}. % perform appropriate transformations for each datatype as part of their data preprocessing step  which can be used to uniformly represent all data fields that need to be visualized in a numerical domain. 
\vizai can be adapted to incorporate these preprocessing steps as well. With numerical data, commonly used  visualizations are line, scatter, area, bar and pie charts, and more sophisticated settings require \emph{density} plots.}

% -- summary of key contributions we make in this paper
Apart from developing a visualization recommendation framework, our work also goes towards developing a systematic performance evaluation measure of various visualizations. In fact, it is an important topic among visualization researchers~\cite{DBLP:journals/cgf/BehrischBKSEFSD18}. We believe that our approach of extracting statistical features of the underlying data mimics the human perception of these visualizations, and offers a framework that can be extended to take into account aspects like choice of colorschemes for the plots and so on.

%%%% - present these later
% In order to confirm if \vizai is able to approximate user's way of interpreting the data in order to visualize it, we conducted a crowdsourced evaluation.
% Through the user survey, we verify that our model mimics user's way of interpreting the data. We did an in depth analysis of different plot types about how good they are at depicting various statistics. With our model one can not only create the best visualization but can also evaluate different plots' performances. We make a contribution towards visualization evaluation which is a hot topic among visualization researchers.
%%%% - 

\subsection{Contributions}
In summary, we make the following key contributions in this paper:

\begin{asparaenum}
\item {We present a novel framework for automatically evaluating the ``goodness'' of a visualization in terms of the accuracy of extracting various statistics of the underlying data, just from the visualization alone. Our approach does not rely on axis or key labels in the plot, or any other numerical/textual content that may be embedded in the plot.}
\item {Based on the above, we present \vizai, an intelligent visualization recommendation system based on generative-discriminative framework. It uses a deep convolutional neural network for generating various statistical features from the plot, and a discriminator that selects the plot with the lowest error in predicting aggregate statistics of the underlying data.}
\item {We conduct a detailed evaluation, using real-world (from Plot.ly) visualizations along with a visualization learning and benchmarking repository, VizNet \cite{viznet} and the results show that \vizai offers high quality recommendations nearly as good as human evaluators.
%\textcolor{blue}{\item To Be Removed - We also perform benchmarking experiments against existing ML-based and rule-based visualization recommendation systems.} 
}
\end{asparaenum}

\subsection{Organization}
% *We need to write the intro and include an interesting example of our model.*
The rest of the paper is organized as follows: in Section~\ref{sec:relwork}, we discuss work related to the ideas pursued in this paper. These range from studies on human perception of visualizations, machine perception of visualizations, and various automated visualization recommendation systems. Next, in Section~\ref{sec:problem}, we present our key hypothesis regarding how to evaluate the quality of visualizations, and formulate our problem statement. Various aspects of \vizai system, including the statistical features and the CNN for extracting these features from visualizations, are described in detail in Section~\ref{sec:vizai}. It is followed by the details of our experimental framework in Section~\ref{sec:crowd}, and the results over real-world datasets in Section~\ref{sec:expresults}. We finally conclude and outline future directions of research in Section~\ref{sec:conclusions}.

\section{Related Work}
\label{sec:relwork}
%We can write about 1. VizML 2. DeepEye 3. Data2Vis 4. Draco-Learn 5. 
Our work focuses on utilizing deep convolutional neural networks trained to emulate human perception of visualizations for automatic selection of best visualization.  In this section we review prior work in each of these areas.

\subsection{Human Perception of Visualizations} 

Early works experimentally demonstrated that aggregate statistics of data, means and correlations respectively could be perceived with high accuracy and compared across classes by humans from data visualizations~\cite{Gleicher2013PerceptionOA, Harrison2014RankingVO}. 
Subsequent analysis of chart perception through extensive human studies found that both individual features as well as aggregate of data could be inferred from good visualizations ~\cite{Kim2018AssessingEO}. They varied the  number of columns and the entropy of data, showing that for a small number of columns humans can perceive differences and even individual values of aggregates like averages with low error rates.

Using a large-scale crowdsourced study, \cite{Saket2018TaskBasedEO} demonstrated that human performance (accuracy as well as time taken) in tasks like finding aggregates (sum for example) of data as well as extremums is highly affected by the type of visualization (scatter, bar, table etc). 

Our work builds on these insights to model the perception of aggregates and extremums to help in identifying the most appropriate visualization form for the given data.

\subsection{CNNs for Data Visualization Understanding}
Deep CNNs are extremely popular for various learning tasks over images~\cite{Ren2015FasterRT,He2016DeepRL}. Considering data visualizations as images, they have been applied for tasks such as visual QA, focus tracking etc. For example, \cite{Kafle2018DVQAUD} use a combination of CNN and OCR output for question answering on visualizations. As a solution to the problem of automatically learning visual importance i.e., the places in an infographic a user is likely to focus on, \cite{Bylinskii2017LearningVI} used a fully convolutional network for creating importance maps.

Recently \cite{Haehn2018EvaluatingP} did an evaluation of CNNs as applied to visualizations. They tried to regress elementary quantitative features like position, length and area from the images of visualizations presented to the network. They suggest that though CNNs can regress quantities for well constrained problems, they would require extensive training to generalize across tasks. 
% For our tasks since the regressed features correspond to 

\subsection{Visualization recommendation} \label{subsec:rec_survey}
There have been many recent efforts towards automating the visualization recommendation process for a given dataset. We briefly survey four of the most recent  works that have shown significant promise in this direction. 

\begin{asparadesc}
\item[Data2Vis ~\cite{Dibia2018Data2VisAG}] {tackles the  visualization generation problem as a language translation problem where data specifications are mapped to visualization specifications. It uses an LSTM based neural translation model to learn a mapping between JSON encoded data and a Vega-lite visualization specification. The model is trained using $4,300$ Vega-lite examples. Their model was in an elementary way able to learn the vocabulary and syntax for a valid visualization specification, appropriate transformations (count, bins, mean) and  how to use common data selection patterns that occur within data visualizations.}

\item[VizML~\cite{vizml}] {formulates visualization recommendation as making design choices. It learns these choices from a large corpus of one million dataset-visualization pairs collected from Plot.ly public studio of visualizations. They use $841$ hand-coded features to train their recommendation networks to make encoding-level and visualization-level design choices. VizML verified that certain steps in a visualization generation pipeline can be learned well using neural networks.}

\item [DeepEye ~\cite{Luo2018DeepEyeTA}] {combines rule based visualization generation with models trained to 1) classify a visualization `good' or `bad' using a decision tree classifier and 2) rank lists of visualizations using a ranking neural network. The DeepEye corpus consists visualizations drawn from 42 public datasets. 100 students  annotated these visualizations as good/bad. These annotations, combined with 14 features for each column pair and rule based heuristics are used to train their models.}
% We can write about 1. VizML 2. DeepEye 3. Data2Vis 4. Draco-Learn 5. 

\item[Draco \cite{Moritz2018FormalizingVD}] {tackles the visualization recommendation in a more formal framework. It uses Answer Set Programming (ASP), a contraint logic programming to specify and generate visualizations with soft weights for constraints learnt from training data.}
\end{asparadesc}

%\textcolor{blue}{To be Removed - Amongst these, only VizML makes their code and data publicly available for conducting empirical comparisons. We use them along with a subset from VizNet~\cite{viznet}, and our own dataset crawled from Plot.ly public website.}

\section{Problem Formulation}
\label{sec:problem}
The visualization of a dataset is a collection of visual elements used to communicate the data in a distortion-free manner. While there are many aspects of visually communicating the contents of a dataset, including the choice of colors, domain-specific imagery such as maps, overlays, etc., our work is based on a fairly simple and fundamental premise that  one should be able to derive the core statistical features of the underlying dataset accurately from the visualization~\cite{tufte2001,tukey1977}. %In this work, we do not focus on artistic elements --such as the use of colors/shades. 
%Data visualization is a collection of visual elements to communicate information of the underlying data with visual elements. 
%We introduce an automated system to determine the best plot given any dataset. 
Since finding the best visualization is a normative question, we make the following assumptions and later validate them through user studies:
\begin{itemize}
    \item Humans prefer visualizations from which they can perceive features about the underlying data accurately.
    \item The modern neural network based models, trained to predict data features from plots, naturally emulate human perception to some extent. That is, if the model makes larger errors in estimating statistical features from one visualization over from another visualization of the same data, we expect that humans when asked to perceive the same statistics end up making corresponding scale of errors\footnote{It should be noted that we \emph{do not} expect the same exact amount of statistical errors between human and our model, but rather their relative scale.}.
\end{itemize}
With these assumptions our problem of finding the best visualization is reduced to the task of designing and training a model which satisfies the above criteria and finding a set of features.

\section{\vizai Framework}
\label{sec:vizai}
\begin{figure*}[t]
    \centering
    \includegraphics[width = 0.7\textwidth]{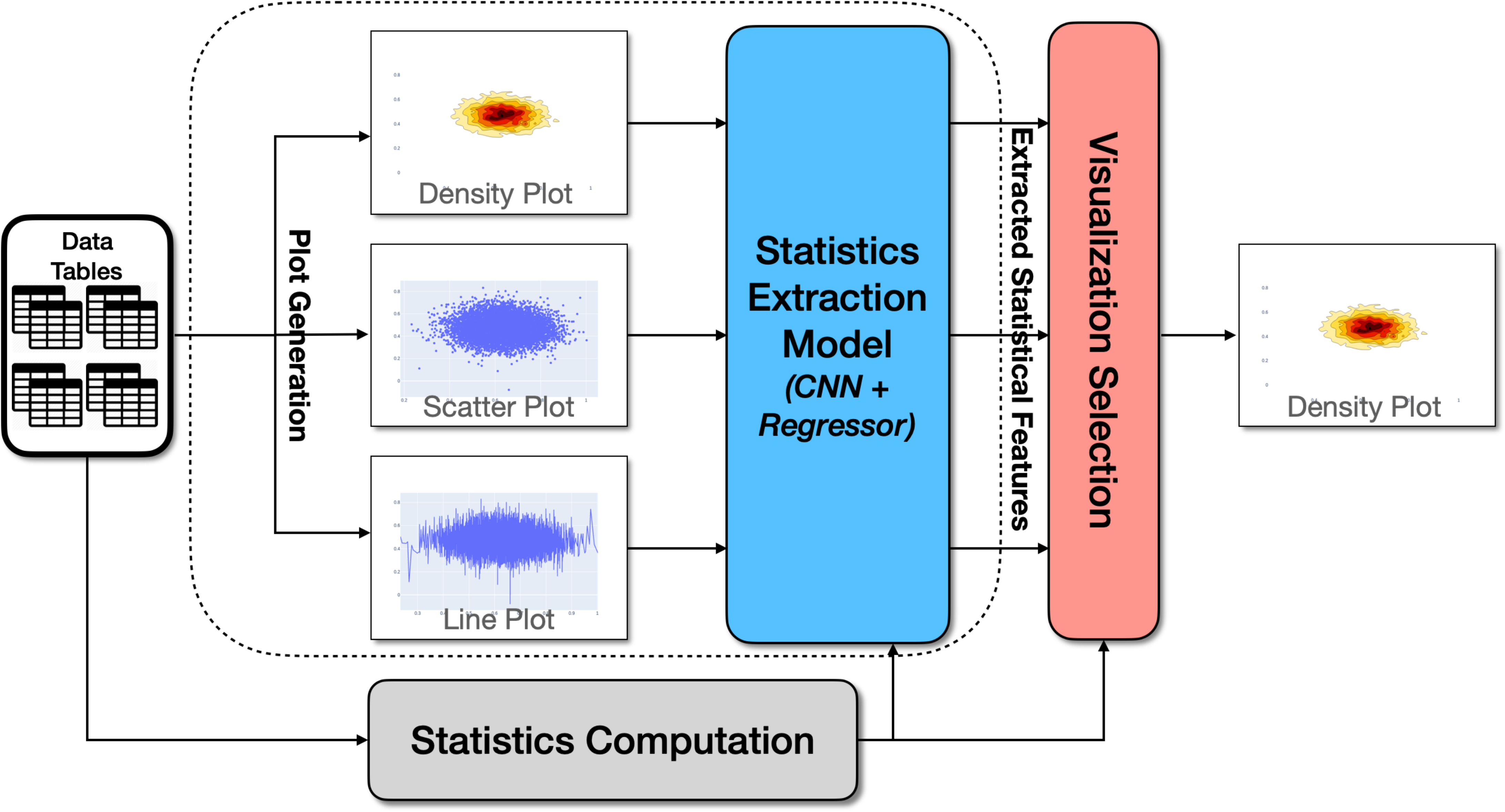}
    \caption{Architecture of the Proposed \vizai Model. The components within the dotted box are trained end-to-end -- note that \vizai does not rely on human labeled data during model training.}
    \label{fig:my_label}
\end{figure*}

\vizai consists of the following three phases: 
\begin{enumerate}
    \item \textbf{True Statistics Computation and Candidate Generation:} We compute a set of statistics from the underlying data tables and use them as features that need to be estimated by our model from the given visualizations. We keep these statistics to be easily computable and, more importantly, easily interpretable by our users. Next, we programmatically generate \textit{a number of visualizations} of each data table using a visualization generator such as Plotly, gnuplot, Vegalite etc.
    \item \textbf{Statistical-Feature Regression:} This is the core model building stage where visualizations are input as images to a supervised Convolution Neural Network (CNN) regression model that extracts (regresses) the statistical features from these images, and trains to minimize the loss between the true statistics and the extracted statistical features. Note that during training, we optimize the statistics regression for each visualization type separately. Thus, adding another visualization type has no impact on the models built on earlier visualization types. 
	\item \textbf{Visualization Selection:} Finally, we compute the loss between the true statistics computed in the first stage and the extracted statistics by the model trained the second stage. The visualization that is considered to be performing best in terms of this loss is recommended for the given data table. 
\end{enumerate}
%During inference, we use the above trained model to predict the best plot for any given data table. This is done by running the CNN+regression model on plots of different types and finding the one where it performs `best'. \\
The overall model architecture is depicted in Figure \ref{fig:my_label}. In the rest of the section, we discuss each of these parts of \vizai in detail.
\subsection{True Statistics Computation and Candidate Visualization Generation}
\begin{table}[tb]
\centering
\begin{tabular}{cc}
\toprule
\textbf{Column level aggregation} & \textbf{Table level aggregation}\\
\bottomrule
Min & Min\\
Max & Max\\
Mean & Mean\\
Std & Std\\
Skew & MAD\\
\bottomrule
\end{tabular}
\caption{Aggregation metrics used to get data features. We also use the correlation feature along with this.}
\label{table:1}
\end{table}
We tried several approaches of selecting a set of statistics as features to extract from data tables. We initially experimented with table embeddings\cite{Gentile2017EntityMO} and other neural feature extraction approaches but their major disadvantage was that these features are not understood or interpretable by end users. Since understanding the various aspects of why a particular visualization is good is relevant for data analysts, for us the interpretability is an essential requirement\cite{Lipton2018TheMO}. Thus finally we settled on a set of simple and commonly used features that most people working with data have some understanding of. These are listed in \ref{table:1}. Note that these are consistent with the recommendation of focusing on \emph{extremes}, \emph{medians} and \emph{relationships}~\cite{tukey1977}. Note that these core statistics can be further enriched as required by specific application settings. 

% Another challenge is that tables can have variable number of columns. 

Although the data tables we currently experiment with consist of only two columns to be visualized, in a general setting we will have to deal with tables with variable number columns. 
%\textcolor{blue}{While the tables we are currently experimenting with consist of only 2 columns, in a general scenario we deal with tables with a variable number of columns.} 
If only individual column-level statistics are used as features, then each data table can result in different feature dimensionality.  To overcome this, we introduce the use of 2-stage feature aggregation. First, column wise statistics (mentioned in the first column of \ref{table:1}) are calculated for each column. Then a data-table level aggregation is carried out using the metrics mentioned in the second column of \ref{table:1}. Thus we use two levels of features, e.g., `max of mean' is the maximum of the mean values of each column. Along with this, we also use Pearson's correlation coefficient, $r$, between any two columns X and Y of our table, written as:
\begin{equation*}
  r =
  \frac{ \sum_{i=1}^{n}(x_i-\bar{x})(y_i-\bar{y}) }{%
        \sqrt{\sum_{i=1}^{n}(x_i-\bar{x})^2}\sqrt{\sum_{i=1}^{n}(y_i-\bar{y})^2}},
\end{equation*}
where $x_i$ and $y_i$ are the $i^{th}$ row for each column and $\bar{x}$ and $\bar{y}$ represents the mean of each column. The generated correlation coefficient in unique for each table as we have only 2 columns in current setup of tables in the dataset. This can be easily modified to performing a table level aggregation of the correlation coefficient to accommodate tables with variable number of columns. As a result, in the current setup each data-table is represented using the same 26 statistical features which are used by the next stage of model building.

%Specify what kind of correlation being used%
\noindent\textbf{Visualization Candidate Generation.} Next, we generate a number of different visualizations for each data table and feed them to a convolutional neural network (CNN) layer coupled with a regressor that extracts statistics from each of these visualizations. As we already mentioned we make use of Plot.ly Chart Studio~\footnote{\url{https://plotly.com/chart-studio/}} for generating visualizations of the table to keep our visualizations consistent with the publicly available datasets from VizNet repository (details in Section~\ref{sec:data}). These visualizations are input to the CNN + regressor model during training, and also as candidate visualizations to select from during the test/inference. 

\subsection{Statistical-Feature Regression}
\begin{figure}[tb]
\centering
\includegraphics[width=0.48\columnwidth]{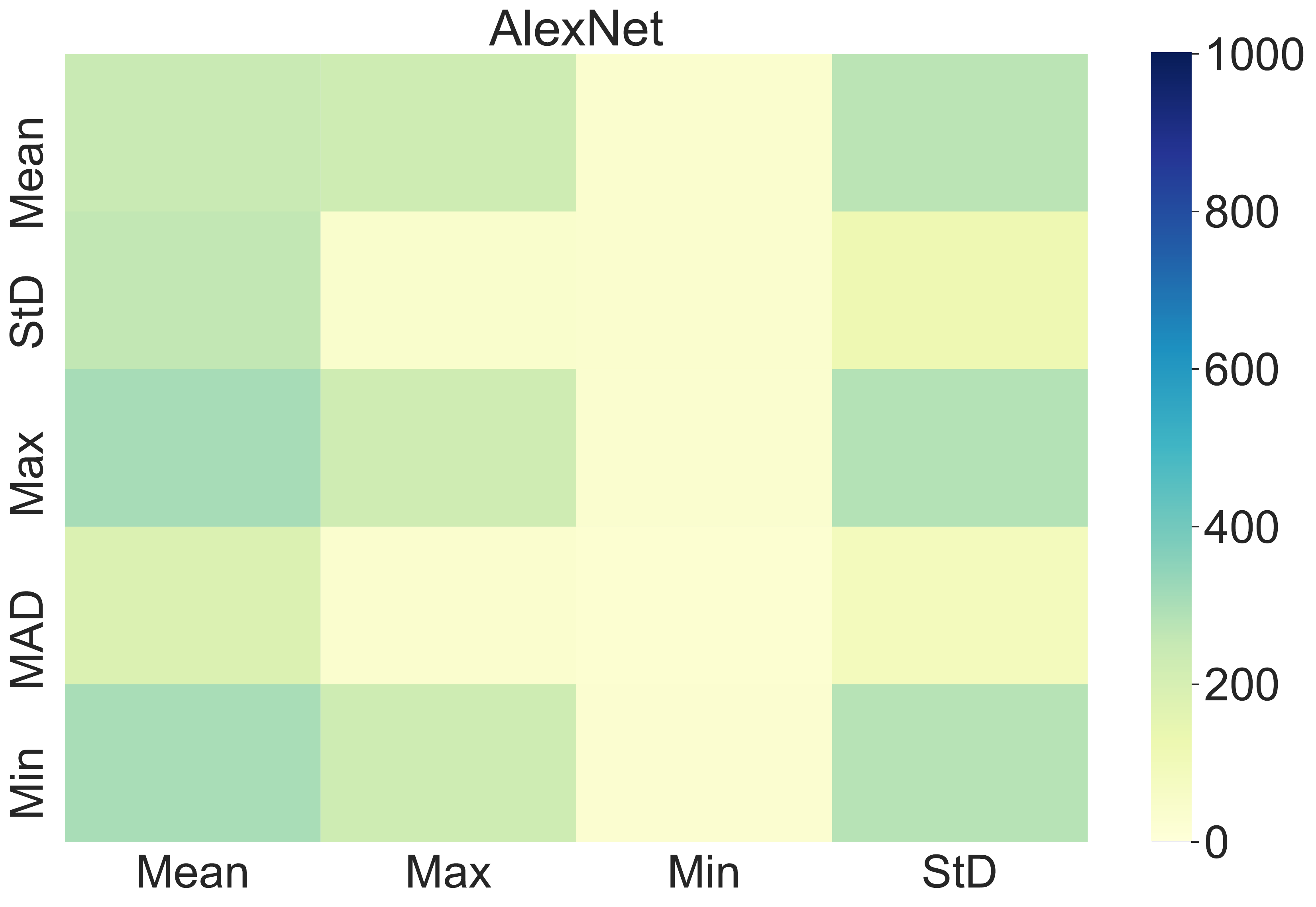}
\includegraphics[width=0.48\columnwidth]{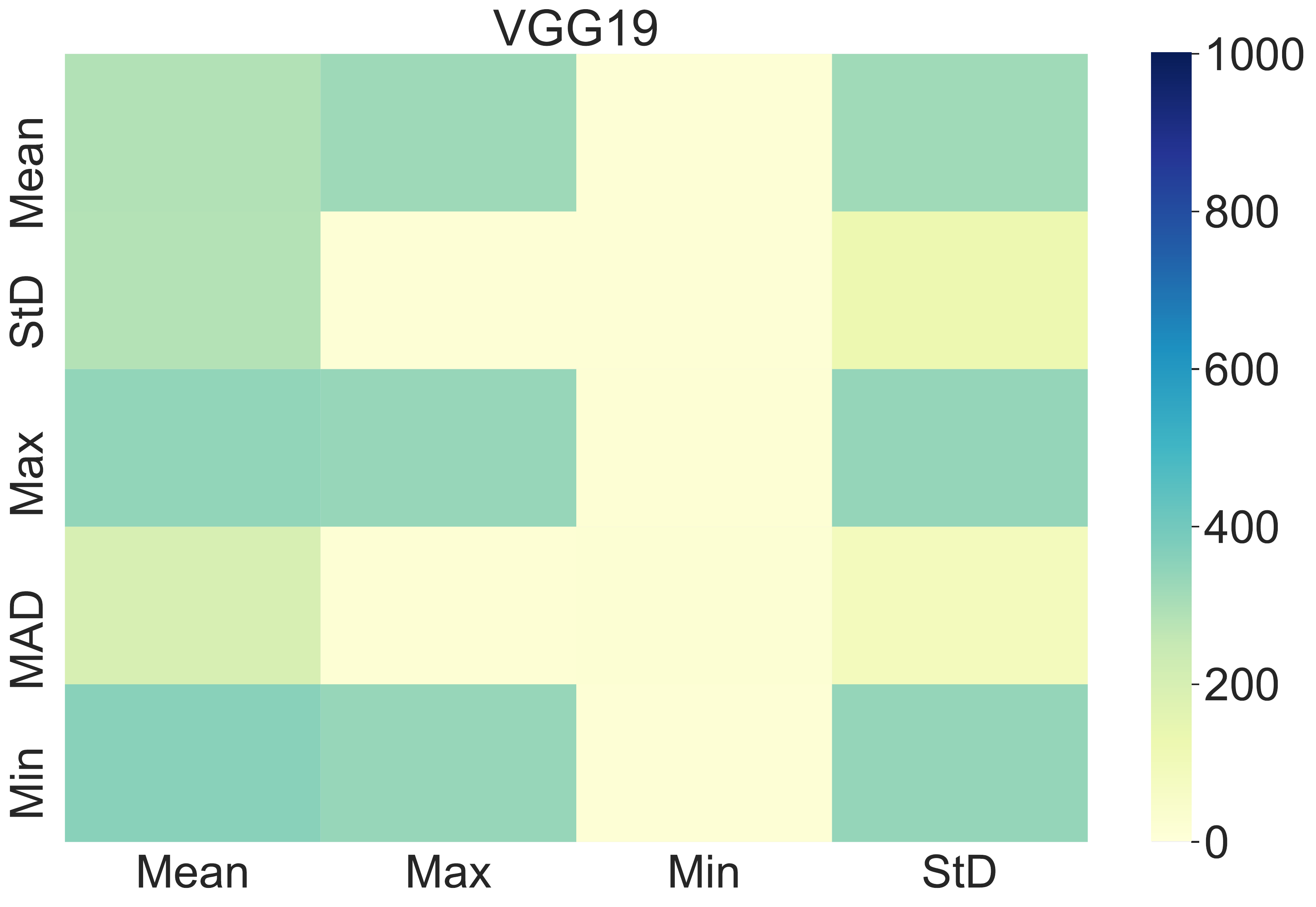}
\includegraphics[width=0.48\columnwidth]{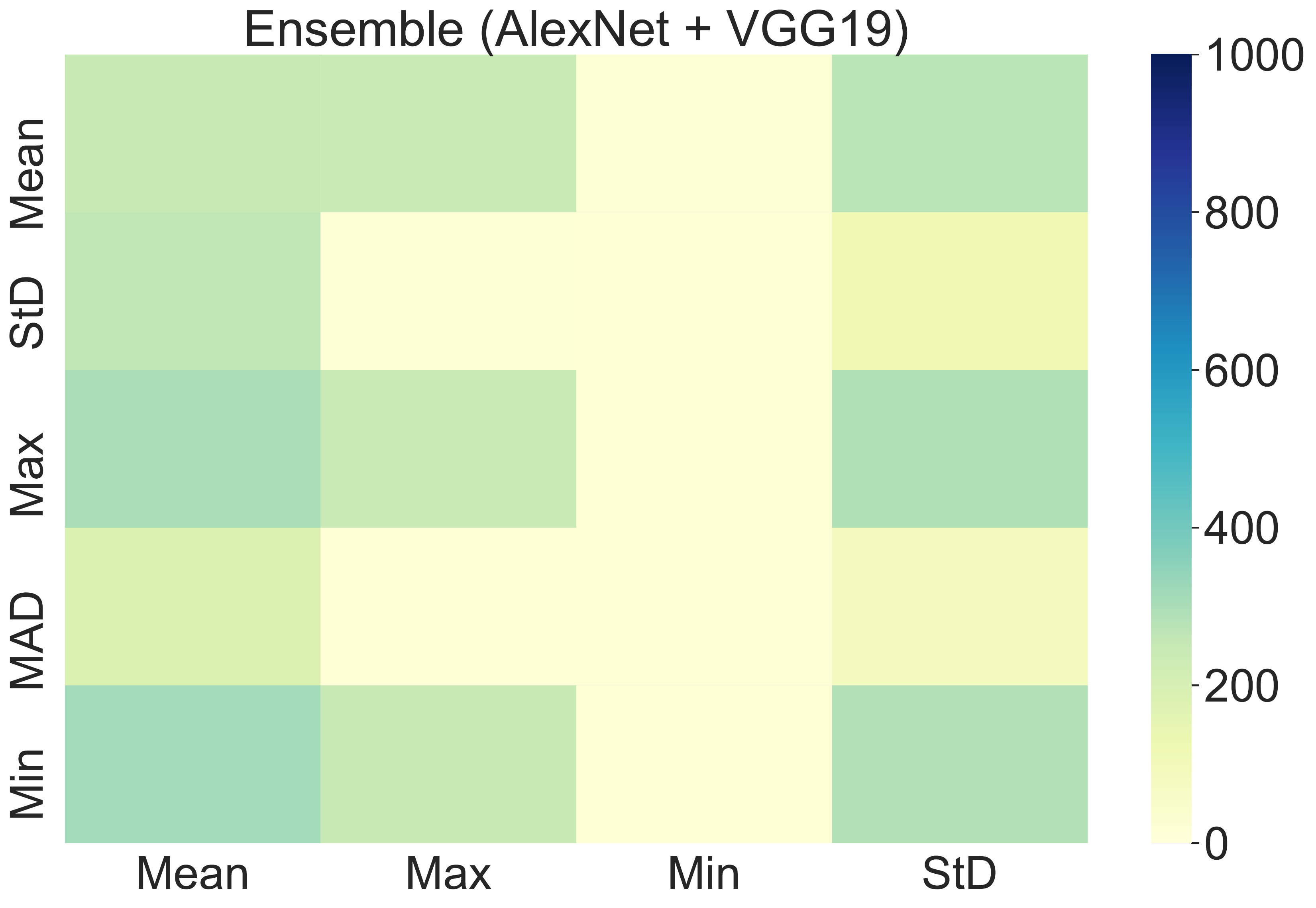}
\includegraphics[width=0.48\columnwidth]{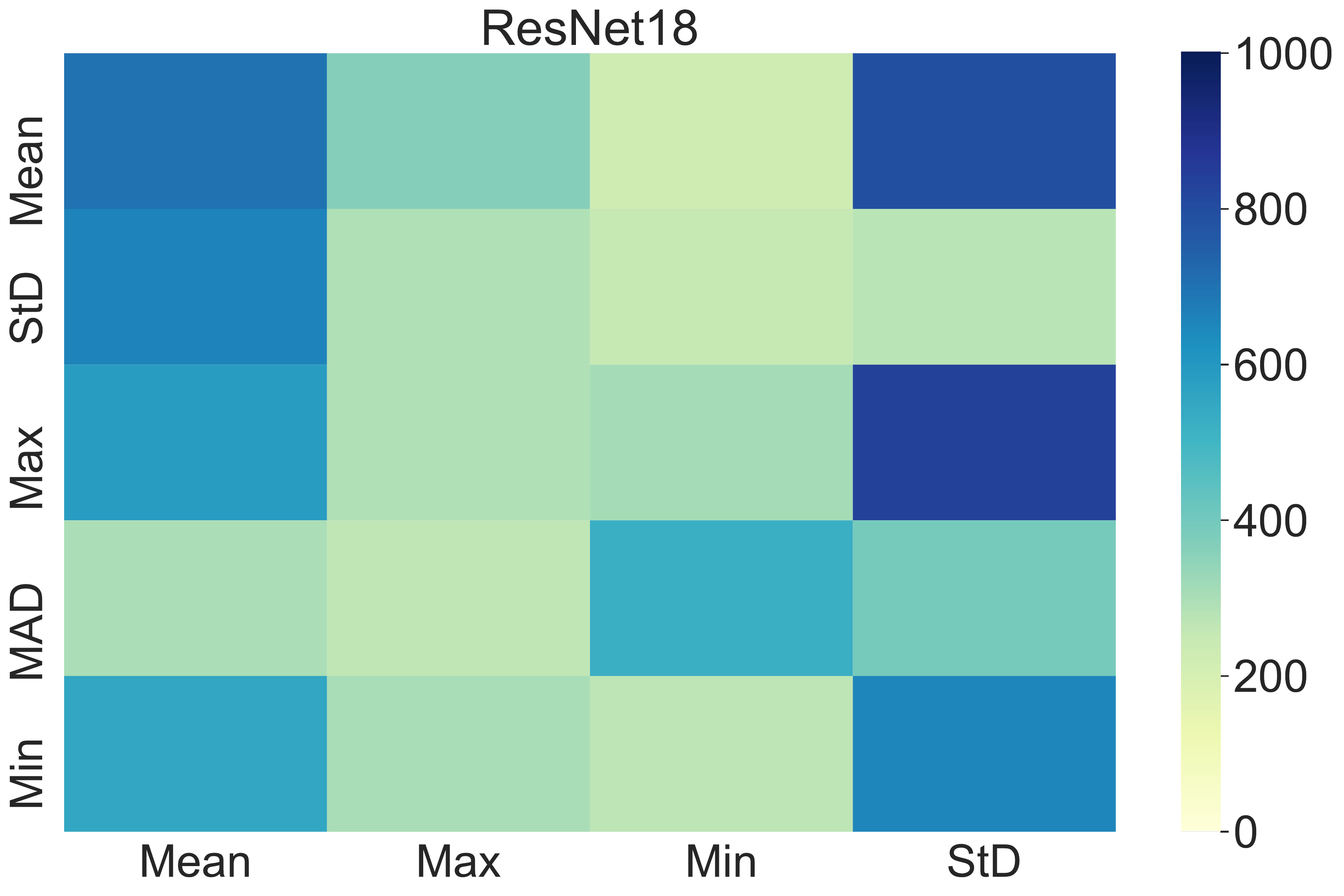}
\includegraphics[width=0.48\columnwidth]{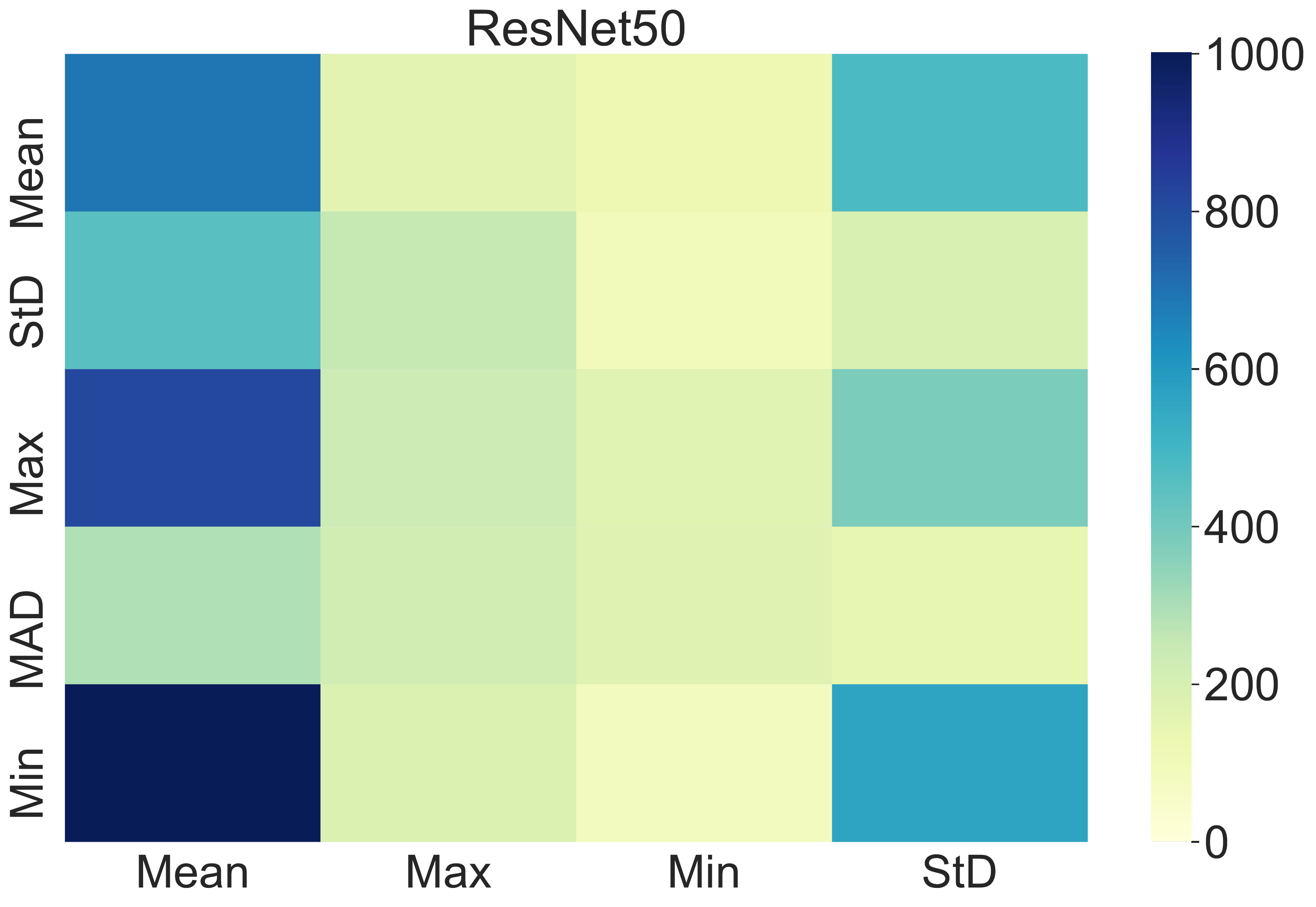}
\caption{Heatmaps illustrating statistics extraction error (loss) on test data of different CNNs trained on Plotly data (darker colors indicate higher error).}
\label{fig:test_loss}
\end{figure}
%
%\includegraphics[scale=0.3]{skew_hm.png}
%\includegraphics[scale=0.3]{corr.png}
%\caption{Plotly Test Dataset loss for each data feature after training different CNN Encoders on Plotly Training Dataset}
%\label{fig:test_loss}
%\end{figure}
 The visualizations are given as image input to the CNN model that generates their embeddings. We train a multivariate regressor to extract various statistical features of the underlying data. 

We experimented with multiple state of the art CNN's in our experiments such as AlexNet \cite{alexnet}, VGG19~\cite{vgg19} along with ResNet18 \cite{He2016DeepRL} and ResNet50 \cite{He2016DeepRL}. As the input images as well as the output labels are quite different from the standard datasets that were used to train these CNNs, simply training the regressor yielded quite poor results. Instead, we simply fine-tune the final classifier layers of CNNs and train the regression layer for our training dataset. 

Somewhat surprisingly we found that larger models (in terms of number of parameters) do not correspond to improved performance in our setting. In  Figure~\ref{fig:test_loss}, we illustrate the performance of different CNNs in terms of their error in extracting various statistics for the dataset we collected from Plotly (details in Section~\ref{sec:data}). It is immediately clear that larger models such as ResNet18 and ResNet50 give huge losses on the test dataset for many statistical features. It may perhaps be due to the issue of over-fitting because of the number of free parameters. On the other hand, AlexNet and VGG19 show consistently better performance across all the statistics we focus on. In our experiments we settled on using \textit{AlexNet} and \textit{VGG19} which were our top-2 best performing CNNs. Additionally, we also experimented with an AdaBoost ensemble of these two models, which we represent as \textit{(AlexNet+VGG19)}.
%
%Talk about the big models and the overfitting in more detail%
%Talk about how imagenet is different from our task%
%

%%%----Not important (Sbj 2021) . Algorithm~\ref{algo:ensemble} lists the way we construct an ensemble model.

% \begin{algorithm}[h!]
%   \caption{The Ensembling Algorithm}
%   \label{algo:ensemble}
%   \textbf{Input:} Training Dataset $P$ with $m$ samples,\\
%     \hspace{0.9cm}  Test Dataset $Q$ with $n$ samples.\\
%   \textbf{Initialisation: $D_{1}[j]=1$ for $j=1,....,m$,\\
% \hspace{2cm}              $\alpha[t]=0$ for $t=1,...,T.$}\\
%   \textbf{Training:}\\
%     \For {$i = 1$ to $T$ trainers} {
%         Train trainer $i$ training dataset $P$.\\
%         Assign integer weights $D_{i}[j]$ for $i=1,....,m.$ after evaluating each $j^{th}$ datapoint on trainer $i$.\\
%         Resample with replacement each $j^{th}$ datapoint $D_{i}[j]$ times in P.\\
%         $\alpha[i] = \frac{1}{2}\log{(\frac{1-err}{err})}$ for trainer $i$ where err is the training loss.
%     }
%   \textbf{Testing:}\\
%     \For{$j = 1$ to $n$}{
%         Prediction $G_j = \frac{\sum_{i=1}^{T}\alpha[i]f_i(Q[j]) }{\sum_{i=1}^{T}\alpha[i]}$ for sample $Q[j]$ and $i^{th}$ trainer $f_i$.
   
%     }
% \end{algorithm}

% \begin{figure}
%     \centering
% \includegraphics[scale=0.31]{lossVSstats.png}
% \caption{This depicts the statistical feature-wise loss on the test dataset after training}
%     \label{fig:overall_model}

% \end{figure}

%need to add captions%

The output embedding from the convolutional layers passes through a fully connected regression layer trained to output the values of statistical features for the given visualization. For the regressor, we used weighted smooth L1 loss between the predicted vector and the features extracted from the data table. The model is trained in a supervised fashion (using the true statistics computed from the previous phase) to predict the statistical features of each visualization.

% \begin{figure}
%     \centering
% \includegraphics[scale=0.4]{reg.png}
% \caption{Overall Model}
%     \label{fig:overall_model}

% \end{figure}

\subsection{Visualization Selection}

At the time of evaluation, we take a data table as our input, extract it's data features for use as true values and produce different types of plot images for it. The regression model is run on each of these to obtain predicted feature values. Based on how close these values are to the actual features of the data, each plot is assigned a score. The best plots is chosen by our model using this score. This not only helps in selecting the best plot for some given data, but also helps in realizing how good a visualization is by analyzing the score. In our experiments, we consider two such types of scoring functions. 

\begin{itemize}
    \item \textbf{Normalised L1 Loss} - For our setup, we define the well known normalised L1 loss as
    % For a data table, we evaluate the predicted statistical features by our CNN Encoder \& Regressor for all the plot types in consideration. 
    % Our discriminator ends up choosing the plot type using the equation 
    
    \begin{equation}
        l^{L1}(\hat{y}, y) = \sum_{i=1}^{26}\frac{|\hat{y_i} - y_i|}{\overline{t_i}}
    \end{equation}
    
    where $\hat{y}$ is the features predicted by our CNN Encoder, $y$ is the true value of all the feature and $\overline{t_i}$ is the average value of the $i^{th}$ feature from the training dataset.
    
    % with the least cumulative L1 loss between predicted statistical features and the true statistical features with normalisation over true statistical features with the average value of each feature from the training dataset.\\
    
    \item \textbf{Top-K Closest Loss} - 
    We define the Top-K Closest loss as the cumulative L1 loss for only the Top-K Statistical features for each plot type prediction closest to true values. 
    
    \begin{equation}
        l^{topK}(\hat{y}, y) = \min {\sum_{i\in tup}(\frac{|\hat{y_i} - y_i|}{\overline{t_i}})} \quad \forall {tup \in \{1,2...,26\}^k}
    \end{equation}
    
    Evidently, this will be nothing but the sum of normalised L1 Loss for the k features whose prediction is closest to true feature values relative to the true value.

\end{itemize}

We will use Normalised L1 Loss, Top-5 and Top-10 Closest Loss in our experiments.

% The individual feature predictions can also be used to interpret visualization effectiveness in a qualitative way.

% \begin{figure}
%     \centering
%     \includegraphics[scale=0.3]{553_hex.png}
%     \caption{Hex plot}
%     \label{fig:my_label}

% \end{figure}
% *Information about Plotly from vizML*
% For the training of the first part of a model we use a Plot.ly corpus scrapped from their publicly available Plotly Commnunity feed. We crawled plots of different plot types like 'areachart', 'line', 'scatter', 'histogram', 'pie'. Around 8000 plots of each type was collected. These data contain. 4 files, namely csv file for the raw data, 

\subsection{Advantages of \vizai}

Based on the literature survey done in section \ref{subsec:rec_survey}, we identify two major limitations of the existing recommendation systems and discuss how \vizai can solve these limitations :

\begin{itemize}
    \item Rule-based recommendation systems like ~\cite{webdb18,Luo2018DeepEyeTA} have obvious limitations when it comes to adapting to new data tables and new types of visualizations. Also machine learning based models such as ~\cite{Dibia2018Data2VisAG,vizml} require complete re-training of the complete dataset in such a setup. \vizai is highly flexible and only requires additional training on new data to adapt to the new setup.
    \item None of these works provide quantification on how well the recommended plots are able to depict the data they were created on. Based on our previous studies, we know the importance of such a quantification, especially for visualisations of data in the numeric domain. Overcoming this limitation, \vizai tries to generate a score for each visualization type based on it's ability to depict data statistics easily. 
\end{itemize} 

\section{Datasets}
\label{sec:data}
As we mentioned earlier, finding the best visualization for a data table is a normative question, user studies are necessary to evaluate the overall performance of recommendations. For this purpose, we conduct experiments using publicly available Plotly data-visualization pairs and a subset from VizNet collection to carefully examine the performance of various choices of statistical-feature regression models, and the loss models used in the visualization selection. %In our second set of experiments, we use much larger synthetic dataset with different set of visualizations as well. 
We generate  plots using the Chart Studio Python API for all of these datasets to maintain uniformity.
% \subsubsection{User visualization preference}
% \textbf{Plot.ly}~\cite{plotly} is an online service used by a large number of people to generate various types of plots of data. We wrote our own crawler which extracted data-visualization pair from the Plot.ly raw feed. We extracted around 8,000 data-visualization pairs each of the following plot types: (i) areachart, (ii)~candlestick, (iii)~hist, (iv)~line, (v) pie. %However since we focus more on numerical data, we ignored box and scattermapbox. 
% We treat the user provided plot type as a gold label for that particular data (majority vote in case of multiple plots of the same data). 

\subsection{Plotly Dataset}
\textbf{Plot.ly}~\cite{plotly} is an online service used by a large number of people to generate various types of plots of data. We wrote our own crawler which extracted data-visualization pair from the Plot.ly raw feed. We extracted data-visualization pairs of the following plot types: (i) scatter, (ii) line, (iii) density. We filtered out data tables with categorical and text columns. For line and scatter plots with more than one dataseries in the same plot, we separated up to 5 of them and treated them as different data-visualization pairs. 

After the preprocessing, the dataset had a total of 24547 data-visualization pairs, the distribution of the data is shown in Table~\ref{table:plotly_data_distr}. We used min-max normalization within a column to keep the range of x and y in [0,1]. Each visualization type was split in 80:10:10 ratio into training, validation and test respectively. 

As we described in Section~\ref{sec:vizai}, our statistics-extraction models based on VGG19, AlexNet, and their ensemble were trained, validated and tested on this dataset. Figure~\ref{fig:test_loss} shows the statistical feature extraction loss we observed on Plotly test data. Note that we make use of only the data tables in this stage to compute the feature extraction loss. 

\begin{table}[tb]
\centering
\begin{tabular}{llll}
\toprule
\textbf{Plot Type} & \textbf{Train} & \textbf{Val} & \textbf{Test}\\
\bottomrule
scatter & 4493 & 562 & 562\\
line         & 11442 & 1430 & 1430\\
density & 3702 & 463 & 463\\
\bottomrule
\end{tabular}
\caption{Plotly Data Distribution after preprocessing}
\label{table:plotly_data_distr}
\end{table}
\subsection{VizNet} VizNet \cite{viznet} is a large-scale repository of data as used in practice, compiled from the web, open data repositories, and online visualization platforms. VizNet repository contains more than 23 visualization types, and a set of  unique data tables and corresponding visualization to represent the data table -- thus,  the visualization type used is considered the \emph{gold label} for the data table. 
We extract 2,042 data table and visualization pairs from the plot.ly fraction of VizNet to build our test dataset. We extract only those instances where (i) the size of data table (i.e., the number of rows) $\geq 5$, and, (ii) the gold labels are either of scatter or line plots as VizNet does not contain density plots as gold label. For each data table extracted, we generate 3 plots of type line, scatter and density. We conducted several experiments and crowd-sourced judgments on this data to validate the performance of our model.

\section{Crowd Sourcing Framework}
\label{sec:crowd}
\begin{figure}
\centering
    
    \includegraphics[width=\columnwidth]{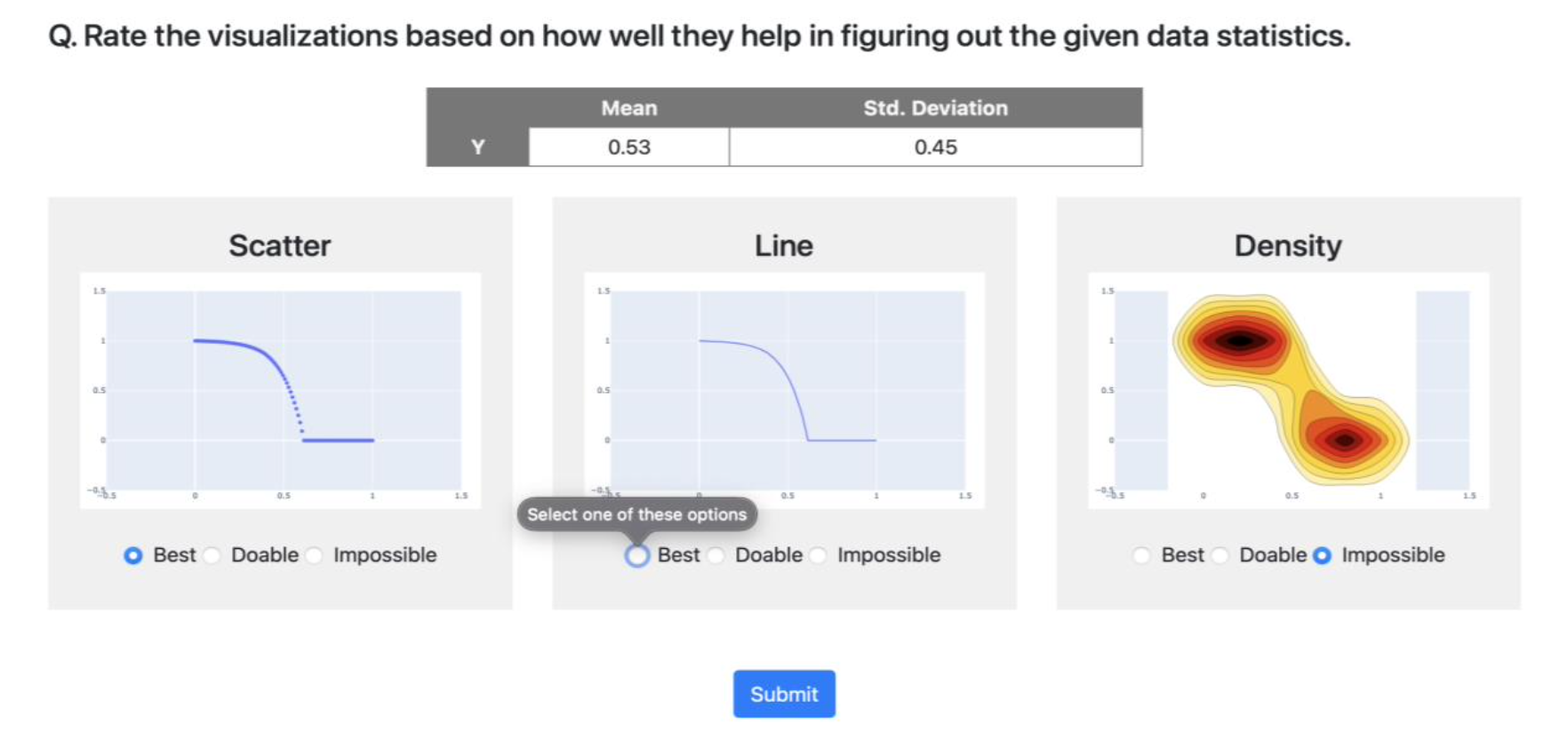}
    \caption{Screenshot of a sample question from the CSI task.}
    \label{fig:CSI}
\end{figure}

For evaluation, we employed a Turkle\cite{hltcoe} instance running on a university server, offering features similar to Amazon Mechanical Turk (AMT) platform. Since the evaluation required expert users who could understand various statistical properties of the data we are interested in inferring, we did not run these experiments directly on AMT. Participants in our evaluation were CS graduate students and faculty members in our university. We defined the following three tasks designed to study our hypothesis about whether statistical features of the underlying data tables are reproduced by the recommended visualizations: %There were four different tasks with each task done by at least 3 different users.
\begin{enumerate}
%     \item \textbf{Simple Statistics Inference (SSI):} Given the minimum and maximum values of X and Y columns of a data table, select the plot which helps identify these extremums the best. All visualization types for the data table are given as choices. The user has to rate on which of the plots it is easiest, doable (with some effort) and impossible for this task. Note that the underlying data table is not given during evaluation. 

    \item \textbf{Complex Statistics Inference (CSI):} Given the mean and standard deviation values of X and Y columns of some data table, the users must select the plot which represents them the best. Similar to the above task, all plots given as choices in every task depict the same data whose statistics are shown to the users (but not the data table itself). The user has to choose on which plots it is easiest, doable and impossible for this task as shown in Figure \ref{fig:CSI}.

    \item \textbf{Fraction Above Threshold (FT):} We provide the users with a single plot per question and ask them to estimate the fraction of points in the given plot which have y (or x) value $> 0.5$. While Answering these questions, It might be trivial to count points for some plots and answer the question so we instructed evaluators to estimate the fraction by looking at the plot without counting the number of points one by one.

%    \item \textbf{Best Visualization (BV):} We presented four different plots representing the same underlying data, and users were asked to select the one which they considered to be visually the best one.
\end{enumerate}

% \begin{figure}
% \centering
%               \includegraphics[width=1\columnwidth]{CSI.png}
%               \caption{Accuracy of visualization prediction by \vizai for different categories over synthetic data.}
%               \label{fig:synth-plotwise-accuracy}
% \end{figure}

\begin{figure}
\centering
    
    \includegraphics[width=\columnwidth]{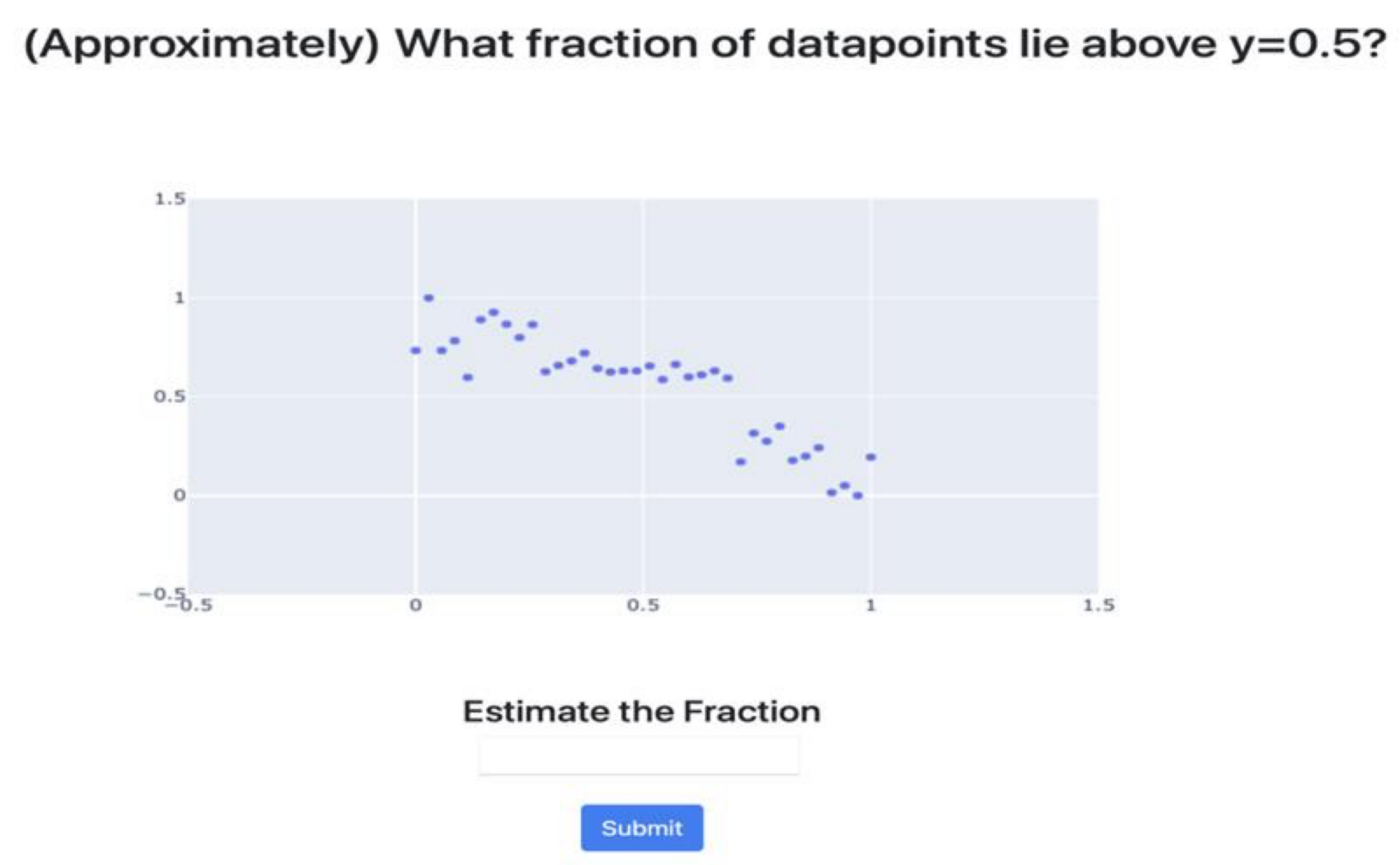}
    \caption{Screenshot of a sample question from the FT task.}
    \label{fig:FT}
\end{figure}

%\sbj{we do not have results for all these questions right? }
%These tasks help us verify the \textit{validity of our hypothesis that the plot that helps realize the underlying statistics of the data correctly is to be preferred.} Using the synthetically generated statistical data as well as statistical data from VizNet \cite{viznet}, we want to identify the plot(s) that help realize the statistics better. Questions 2 and 3 are designed to address this part.%\\ On the other hand, the second part of the hypothesis - whether the plot that the user preferred is indeed the preferred plots for the data is hard to answer using synthetic dataset that we generated. To validate this aspect, we also created three plot types on VizNet data and asked users to select the best plots.
    
%    \item \textit{Whether a model learnt based on this hypothesis generates plots that are “attractive” (visually as well as in terms of their informativeness).}
%        On synthetic datasets, we asked this question after informing the user about the nature of the data we are plotting.
%        On VizNet dataset, we asked more meaningful questions — Given the description of the data, if the plot is appropriate  for the data being described and how easily we can get the idea of important features of the data depicted by the plot.
%\end{enumerate}

We assign points to each plot type for each data table on the basis of answers given by users in the crowdsourcing experiment for VizNet data as shown in Table~\ref{table:crowd_pointsystem}. For CSI task, the points are awarded based on the choice indicating the ease of statistics inference from the plot. In the FT task, where the evaluators have to estimate the fraction, the points are awarded based on the relative error made by them from the true value of the statistic -- upto 20\% error we award 2 points, upto 40\% error we award 1 point, and 0 points are awarded for errors beyond 40\%.  The cumulative points are calculated for each plot type for each data table and the plot type with most number of points is chosen as the overall preferred choice of plot type. 

\begin{table}[tb]
\centering
\begin{tabular}{ccc}
\toprule
\textbf{Points} & \textbf{Chosen Ease of SI} & \textbf{Error in FT Task} \\
\midrule
2 & {Easiest} & $|\hat{y}-y|\leq0.2y$ \\
1 & {Doable} & $|\hat{y}-y|\leq0.4y$ \\
0 & {Impossible} & $|\hat{y}-y|>0.4y$ \\
\bottomrule
\end{tabular}
\caption{Scheme of awarding points to individual data-visualization pairs based on the crowdsourced tasks. For SI tasks, the points are awarded for each user choice. For FT tasks, the points are awarded based on the error in estimation made by users -- $\hat{y}$ represents fraction predicted by user and $y$ represents true fraction. }
\label{table:crowd_pointsystem}
\end{table} 
% \begin{table}[tb]
% \centering
% \begin{tabular}{cccc}
% \toprule
% \textbf{} & \textbf{Best} & \textbf{Doable} & \textbf{Impossible}\\
% \bottomrule
% Points & 2 & 1 & 0\\

% \bottomrule
% \end{tabular}
% \caption{Point System for SSI and CSI tasks}
% \label{table:synth_ssi_csi_points}
% \end{table}

% \begin{table}[h!]
% \centering
% \begin{tabular}{cccc}
% \toprule
% \textbf{} & \textbf{$|\hat{y}-y|\leq0.2y$} & \textbf{$|\hat{y}-y|\leq0.4y$} & \textbf{$|\hat{y}-y|>0.4y$}\\
% \bottomrule
% Points & 2 & 1 & 0\\

% \bottomrule
% \end{tabular}
% \caption{Point System for FT tasks where $\hat{y}$ represents fraction predicted by user and $y$ represents true fraction}
% \label{table:synth_ft_points}
% \end{table}

\section{Experimental Results}
\label{sec:expresults}

\subsection{Statistics Extraction Accuracy}
\begin{figure}[tb]
\centering
\includegraphics[width=0.7\columnwidth]{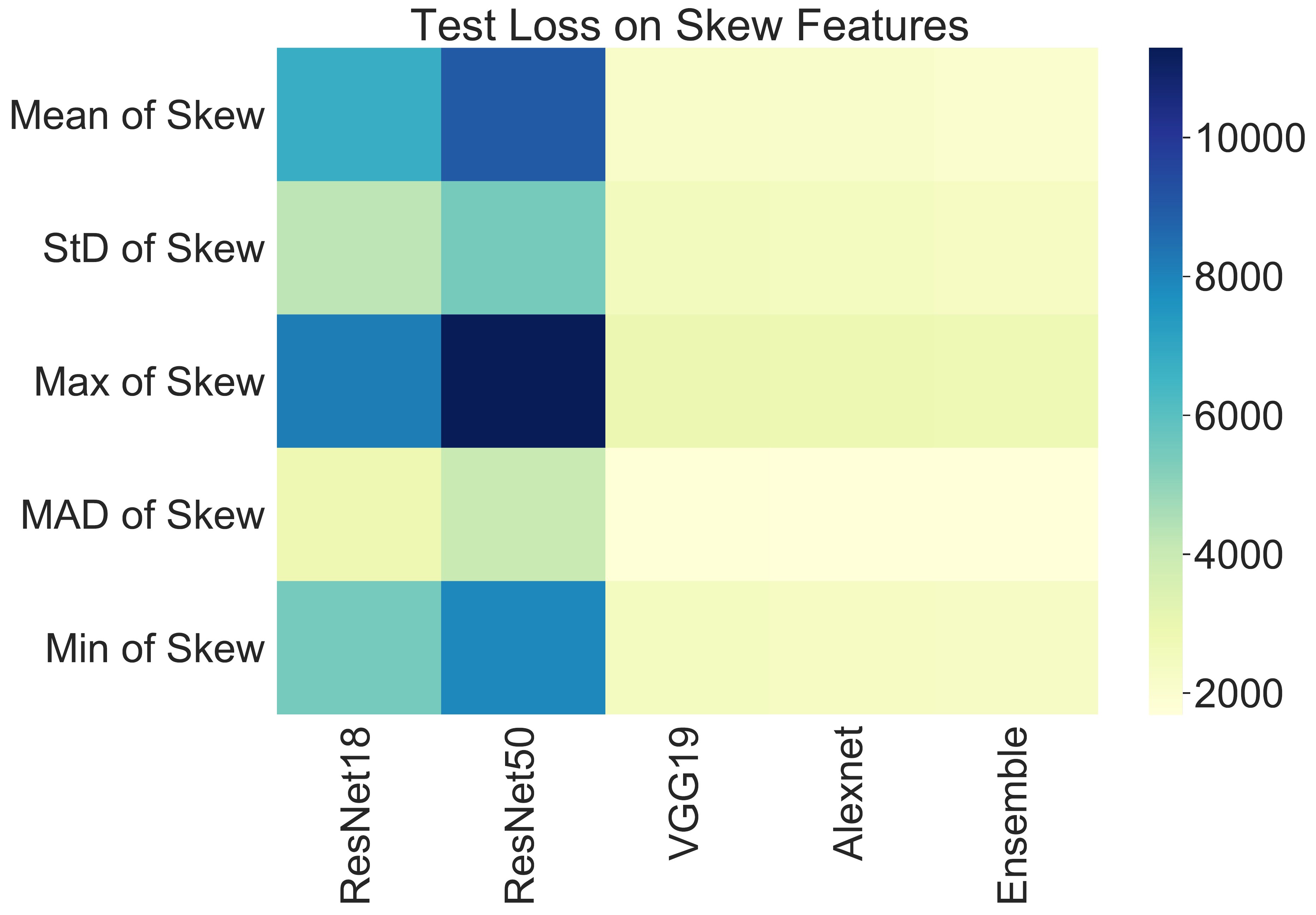}
\caption{Heatmap of loss observed on Skew column aggregate statistic under various  extraction models.}
\label{fig:loss_skew}
\end{figure}

\begin{figure}[tb]
\centering
\includegraphics[width=0.7\columnwidth]{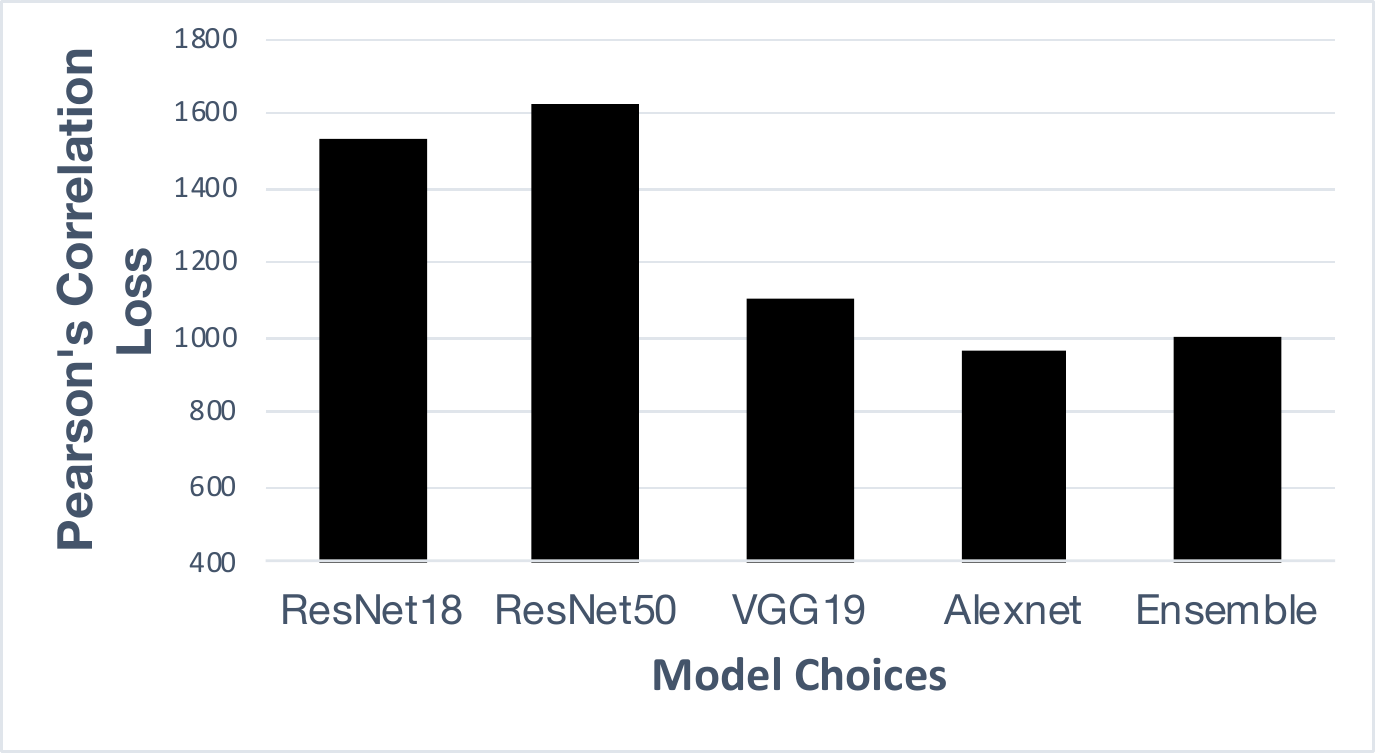}
\caption{Variation in loss in estimating Pearson's correlation between the two columns of the data-table under various models.}
\label{fig:loss_correlation}
\end{figure}

As already mentioned, the statistics extraction model consisting of CNN followed by a fully connected regressor was trained and tested using Plotly dataset.  Figure~\ref{fig:test_loss} presents the heatmaps of statistics extraction error with different choices of CNNs including our AlexNet+VGG19 ensemble. In Figure~\ref{fig:loss_skew} we show the loss in the \emph{Skew} column aggregation over each of the table level aggregations given in Table~\ref{table:1}. Similarly, Figure~\ref{fig:loss_correlation} plots the loss in estimating Pearson's correlation coefficient between X and Y columns of the data table from the visualization alone. As one can observe, the larger CNN models (ResNet18 and ResNet50) perform much poorer than smaller models, and our (VGG19+AlexNet) ensemble further improves on statistics extraction accuracy. 

\subsection{VizAI Performance on VizNet}

\begin{table*}[h!]
\centering
\resizebox{\textwidth}{!}{\begin{tabular}{c|c|ccccc}
\toprule
\textbf{Statistics Extraction Model} & \textbf{Visualization Selection Model} & \textbf{Accuracy} & \textbf{F1-Score(Scatter)} & 
\textbf{F1-Score(Density)} & 
\textbf{F1-Score(Line)} & \textbf{Weighted F1-Score}\\
\midrule
\multirow{3}{*}{AlexNet} & L1-Loss & 0.60 & 0.65 & 0.56 & 0.46 & 0.59\\ 
			 & Top-5 Closest Loss & 0.71 & 0.80 & 0.62 & 0.40 & 0.68\\
			 & Top-10 Closest Loss & 0.69 & 0.76 & 0.59 & 0.46 & 0.66\\
\cmidrule(lr){1-7}                  
\multirow{3}{*}{VGG19} & L1-Loss & 0.54 & 0.50 & 0.54 & 0.50 & 0.51\\
			 & Top-5 Closest Loss & 0.63 & 0.61 & 0.46 & 0.67 & 0.57\\
			 & Top-10 Closest Loss & 0.63 & 0.61 & 0.46 & 0.67 & 0.57\\
\cmidrule(lr){1-7}                  
\multirow{3}{*} {(AlexNet+VGG19)} & L1-Loss & 0.66 & 0.70 & 0.61 & 0.43 & 0.63\\
                & Top-5 Closest Loss & \textbf{0.80} & \textbf{0.86} & \textbf{0.67} & \textbf{0.55} & \textbf{0.75}\\
			    & Top-10 Closest Loss & 0.77 & 0.83 & 0.62 & 0.67 & 0.74\\

\bottomrule
\end{tabular}}
\caption{Performance of different Statistics Extraction and Visualization Selection models on VizNet. The bold-faced values represent the best performing combination.}
\label{tab:viznet_results}
\end{table*}

%\begin{table*}[h!]
%\centering
%\begin{tabular}{cccccc}
%\toprule
%\textbf{Combination} & \textbf{Accuracy} & \textbf{F1-Score(Scatter)} & \textbf{F1-score(Density)} & \textbf{F1-score(Line)} & \textbf{Weighted F1-Score}\\
%\bottomrule
%(1,1) & 0.71 & 0.80 & 0.62 & 0.40 & 0.68\\
%(1,2) & 0.69 & 0.76 & 0.59 & 0.46 & 0.66\\
%(1,3) & 0.60 & 0.65 & 0.56 & 0.46 & 0.59\\
%(2,1) & 0.63 & 0.61 & 0.46 & 0.67 & 0.57\\
%(2,2) & 0.63 & 0.61 & 0.46 & 0.67 & 0.57\\
%(2,3) & 0.54 & 0.50 & 0.54 & 0.50 & 0.51\\
%\rowcolor{lightgray} (3,1) & 0.80 & 0.86 & 0.67 & 0.55 & 0.75\\
%(3,2) & 0.77 & 0.83 & 0.62 & 0.67 & 0.74\\
%(3,3) & 0.66 & 0.70 & 0.61 & 0.43 & 0.63\\
%\bottomrule
%\end{tabular}
%\caption{Each Combination (i,j) represents (Generator, Discriminator) pair where i and j are the ID's from Table 6. The shaded row represents the best performing pair on the VizNet data.}
%\label{table:bigmetrics}
%\end{table*}

%  \label{sec:viznet_results}
% \begin{figure*}[t]
%     \centering
%     \includegraphics[width = \textwidth, height = 160pt]{viznet_table.png}
%     \caption{Each Combination (i,j) represents (Generator, Discriminator) pair where i and j are the ID's from Table 6. The boxed row represents the best performing pair on the VizNet data.}
%     \label{Table 7}
% \end{figure*}
We now turn our attention to evaluating the performance of VizAI on real-world data from VizNet~\cite{viznet} based on a crowd-sourced experiment. We selected samples from VizNet which are 2-Dimensional and have gold labels as either line or scatter plots (note that VizNet does not have density plots). We generated the same three plots for each sample for the crowd sourcing experiment. We removed the labels of the data and just treated the two columns as X and Y.

\begin{table}[h!]
\centering
\begin{tabular}{cc}
\toprule
\textbf{Plot Type} & \textbf{Preferred Plot Type Count}\\
\bottomrule
Scatter & 21\\
Density & 12\\
Line & 6\\

\bottomrule
\end{tabular}
\caption{This table shows the number of times each plot type is chosen as the preferred user plot type after assigning points in the VizNet data}
\label{table:agreement}
\end{table}

We asked each participants in the crowdsourcing to complete a total of 245 SI and FT (both for Y and X axis) tasks, on a randomly selected 35 tables from the test set. We aggregate points based on the scheme described earlier (Table~\ref{table:crowd_pointsystem}) for CSI and FT tasks. The plot type with maximum points is chosen as the preferred plot type. The resulting count of preferred plot type for each visualization type is summarized in Table~\ref{table:agreement}. Note that the sum of these counts is greater than 35 due to ties in the points between two or more plot types. It is interesting to observe that although VizNet did not have density plots, there were a significant instances where crowd evaluators shown preference to density plots over line chart. In our subsequent evaluation, we used our crowd sourced plot types as gold labels. 

% We quantify responses by using separate point systems for CSI tasks (Table~\ref{table:synth_ssi_csi_points}) and for FT tasks (Table~\ref{table:synth_ft_points}). The plot type with most number of points is chosen as the user's preferred choice of plot type.

% \begin{table}[h!]
% \centering
% \begin{tabular}{ccc}
% \toprule
% \textbf{ID} & \textbf{Generator} & \textbf{Discriminator}\\
% \bottomrule
% 1 & AlexNet & L1 Loss\\
% 2 & VGG19 & Top-5 Closest Loss\\
% 3 & Ensemble(AlexNet + VGG19) & Top-10 Closest Loss\\

% \bottomrule
% \end{tabular}
% \caption{This table represents different generators and discriminators we will experiment with.}
% \label{table:match}
% \end{table}

We experiment using all combinations of statistics extraction models (AlexNet, VGG19 or (VGG19+AlexNet) ensemble) and the visualization selection (L1-loss, Top-5 Closest and Top-10 Closest loss) models. The results are summarized in Table~\ref{tab:viznet_results}. From these results, we can evidently see that the use of the (VGG19+AlexNet) ensemble, followed by the use of Top-5 Closest loss in the visualization selection performs the best in terms of accuracy and F1-scores of individual plot types as well as aggregated weighted F1-score using the counts from Table~\ref{table:agreement}. 

%and weighted F1-score. This pair achieves 80\% accuracy and 0.75 weighted F1-score with weights being the Human Preference Count for each plot type from Table~\ref{table:agreement}.
%The performance on individual plot types for the winning generator-discriminator pair are given in Table~\ref{table:breakup}.

% \begin{table}[h!]
% \centering
% \begin{tabular}{cccc}
% \toprule
% \textbf{} & \textbf{Precision} & \textbf{Recall} & \textbf{F1-Score}\\
% \bottomrule
% Scatter & 1.0 & 0.76 & 0.86\\
% Density & 0.6 & 0.75 & 0.67\\
% Line & 0.6 & 0.5 & 0.55\\

% \bottomrule
% \end{tabular}
% \caption{Evaluation Metrics for pair ((Ensemble Model, Top-5 Closest Loss)}
% \label{table:breakup}
% \end{table} 
\subsection{Issues with VizNet Gold Labels}
\begin{figure}[t]
\begin{center}
   \includegraphics[width=1.\linewidth]{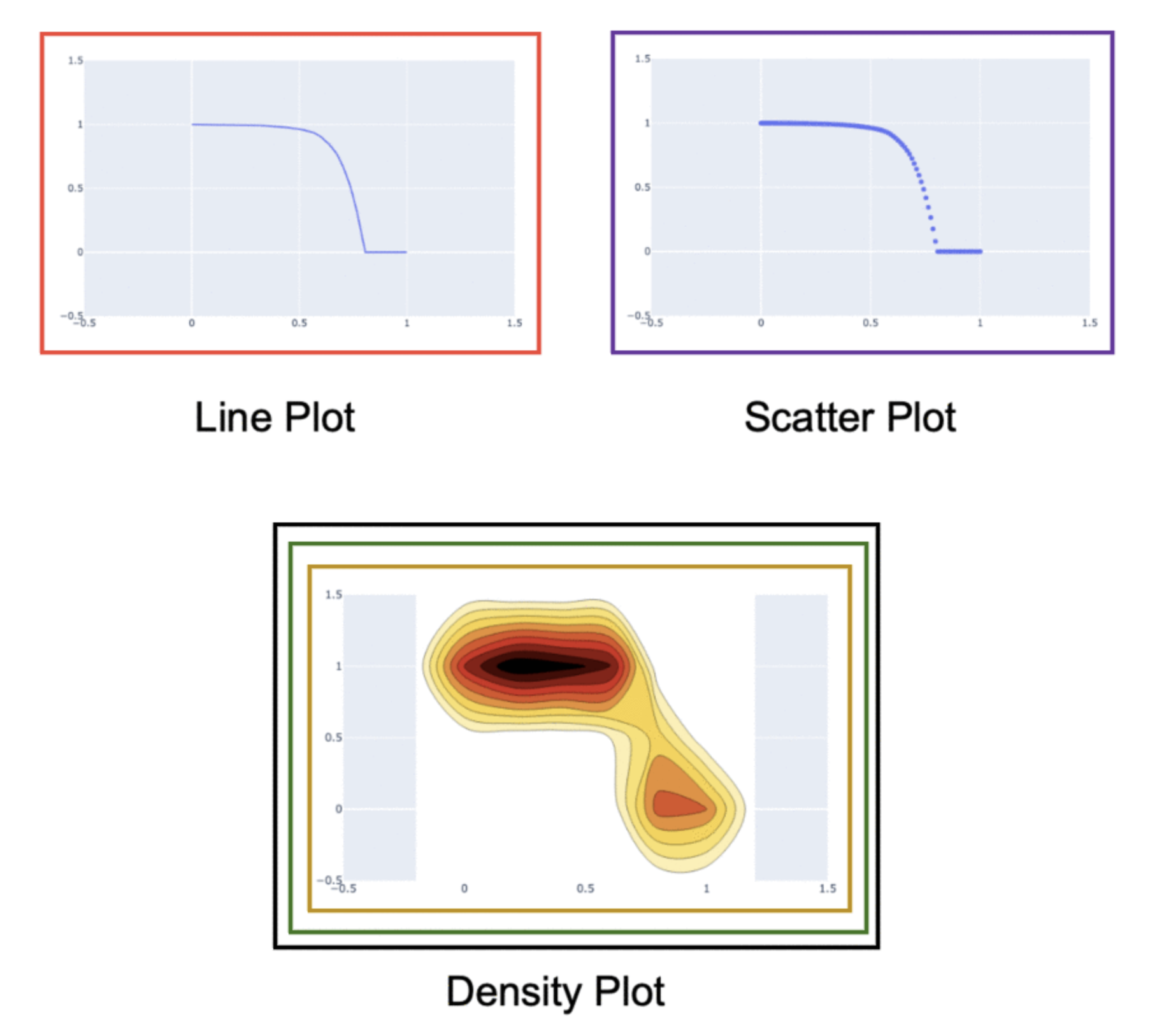}
\end{center}
\caption{VizNet Gold Label choice is highlighted in red box, VizAI's choice is highlighted in the black box and the crowdsourcing users 1,2 and 3's choice are highlighted in the blue, yellow and green boxes respectively. VizNet Gold Label choice is unreliable as the line plot gives us no idea about the distribution of the points and hence the underlying statistics of the data.}
\label{fig:vizai_vs_viznet}
\end{figure}

As we observed deviations in crowdsourced gold labels we collected from the gold labels  given in VizNet dataset, we felt it prudent to investigate this issue further. We first evaluate our best performing combination -- (VGG19+AlexNet) with Top-5 closest loss -- when using the gold labels as given in VizNet. Table~\ref{table:match} shows that the use of VizNet gold label results in much worse performance. However, we believe that this is indicative of the data quality issues one would expect when crawling at large (as observed recently in~\cite{stochasticparrots}). We further highlight this issue with an example drawn from our crowdsourcing experiment shown in Figure~\ref{fig:vizai_vs_viznet}. While VizNet gold label was the line plot, two evaluators for this task clearly preferred the density plot (and one evaluator chose the scatter plot).

% We find that the predicted plots by VizAI are far better than the plot type labels given in the original VizNet dataset in the sense that more number of users agreed on them being the best visualization. One such example is shown in figure \ref{fig:vizai_vs_viznet}.

\begin{table}[tb]
\centering
\begin{tabular}{ccc}
\toprule
\textbf{Metrics} & \textbf{VizAI} & \textbf{VizNet Gold Label}\\
\bottomrule
Accuracy & 0.80 & 0.43\\
F1-Score(Scatter) & 0.86 & 0.59\\
F1-Score(Line) & 0.55 & 0.29\\
\bottomrule
\end{tabular}
\caption{Performance comparison of VizAI vs plot type labels given in the original VizNet dataset}
\label{table:match}
\end{table}

\section{Conclusions}
\label{sec:conclusions}
In this work we presented VizAI, a visualization recommendation system  based upon machine prediction of statistics of the data from different plot images. 
%\textcolor{blue}{To Remove - Our experiments demonstrate that our model is better than state of the art approaches at predicting visualizations favored by end users.}
Our experiments demonstrate that our model makes intelligent visualization predictions which are highly favored by end users. The results of our crowd-sourced human study reveals interesting aspects behind human perception of visualizations. 

From a practical perspective, the approach taken by VizAI can easily adapt to new visualization paradigms without significant cost compared to VizML which requires complete retraining to accommodate for the new plot type. Further, it is possible to incorporate domain-specific details like focus on the quality of certain statistics, color preferences, different visualization libraries, etc., into VizAI.  

\subsection{Limitations}
A limitation of our work is that we make use of an algorithmic system, viz., deep convolutional neural network (refer Section~\ref{sec:vizai}), as a proxy for human visual perception to determine the extent to which a visualization is representing various data statistics. However, it is well-known in visualization literature that human visual perception is highly context dependent (e.g., it is subject to optical illusions), and  varies greatly from person to person. Naturally, CNNs are not expected to fully mimic the human perception of visualizations. We believe that further research is required, especially in complex visualizations of semantically rich data. Despite this, \vizai offers a unique and flexible approach to recommending visualizations for a dataset. 

\subsection{Future Work}
We plan to pursue research in extending \vizai along following three directions:
\begin{enumerate}
    \item  Incorporation of more plot variables such as - which columns to plot, what goes on which axis and more plot types into our prediction framework. For simplicity and easy testing, we chose only to predict plot types but for usability purposes it would be desirable to have multiple decision variables inferred automatically.
    \item Applications of our model, which predicts human perceived data features from plot images  and compares them with actual data features, in different tasks like identifying misleading visualizations (for detecting fake news for example) and helping analysts to interpret plot visualizations. 
    \item More robust scale and data statistic invariance in our feature prediction model. Right now we use a simple normalization procedure to deal with different scales in our input data features. An approach based on learning scale invariant models should improve performance and make the model more adaptable. \end{enumerate}{}

%\newpage

\bibliographystyle{ACM-Reference-Format}
\bibliography{ijcai19}

\end{document}